\documentstyle[12pt,amssymb,amsmath]{article}
\newtheorem{theorem}{Theorem}[section]

\newcommand{\p}{\partial}
\newcommand{\la}{\lambda}
\newcommand{\s}{\sum_{i=1}^N}

\begin{document}

\title{ {\bf The constrained dispersionless mKP hierarchy and the dispersionless mKP hierarchy with self-consistent
sources} }
\author{ {\bf  Ting Xiao
    \hspace{1cm} Yunbo Zeng\dag } \\
    {\small {\it
    Department of Mathematical Sciences, Tsinghua University,
    Beijing 100084, China}} \\
    {\small {\it \dag
     Email: yzeng@math.tsinghua.edu.cn}}}

\date{}
\maketitle
\renewcommand{\theequation}{\arabic{section}.\arabic{equation}}

\begin{abstract}
We first show that the quasiclassical limit of the squared
eigenfunction symmetry constraint of the Sato operator for the mKP
hierarchy leads to a reduction of the Sato function for the
dispersionless mKP hierarchy. The constrained dispersionless mKP
hierarchy (cdmKPH) is obtained and it is shown that the
(2+1)-dimensional dispersionless mKP hierarchy is decomposed to
two (1+1)-dimensional hierarchies of hydrodynamical type. The
dispersionless mKP hierarchy with self-consistent sources
(dmKPHSCS) together with its associated conservation equations are
also constructed. Some solutions of dmKPESCS are obtained by
hodograph reduction method.
\end{abstract}

\hskip\parindent


\section{Introduction}
\setcounter{equation}{0} \hskip\parindent Recently much attention
has been focused on the study of dispersionless integrable
hierarchies which have important applications from complex
analysis to topological field theory (see [1-16]). In
\cite{Lebedev1979,Zakharov1980,Kodama1988,Kodama1989,Takasaki1995},
a standard procedure of dispersionless limit of integrable
dispersionfull hierarchies is proposed. In this procedure,
dispersionless hierarchies arise as the quasiclassical limit of
the original dispersionfull Lax equations performed by replacing
operators by phase space functions, commutators by Poisson
brackets and the role of Lax pair equations by conservation
equations (or equations of Hamilton-Jacobi type ). Several
strategies have been proposed to investigate these hierarchies,
such as the twistorial method
\cite{Takasaki1992,Takasaki1995,Martinez2003}, the hodograph
reduction method \cite{Kodama1988,Kodama1989,Manas2002} and the
quasi-classical $\bar{\partial}$-method [13-16]. Lately in
\cite{Bogdanov2003}, using the quasi-classical
$\bar{\partial}$-method, the authors studied the symmetry
constraint of the dispersionless KP (dKP) and 2DTL hierarchies and
some important reductions of these hierarchies are shown to be
nothing but results of symmetry constraints. This offers us
another important way to construct reductions of dispersionless
hierarchies.

The integrable equations with self-consistent sources are another
type of important integrable models in many fields of physics,
such as hydrodynamics, state physics, plasma physics, etc (see
[19-27]). In our general approach proposed recently in [22-27],
the constrained integrable hierarchy may be regarded as the
stationary system of the corresponding integrable hierarchies with
self-consistent sources. From this observation, the integrable
hierarchies with self-consistent sources and their Lax
representation can be constructed from the corresponding
constrained integrable hierarchies. The KP hierarchy with
self-consistent sources (KPHSCS) and the modified KP hierarchy
with self-consistent sources (mKPHSCS) are constructed in this way
and their interesting solutions are obtained via generalized
Darboux transformation \cite{Xiao20041,Xiao20042}. In this sense,
the soliton hierarchies with self-consistent sources may be viewed
as integrable generalizations of the original soliton hierarchies.

In \cite{Xiao2005}, at the first time we studied the
quasiclassical limit of the symmetry constraint of the Sato
operator for the KP hierarchy in the framework of Sato theory
which offers us a new way to study the reduction of the
dispersionless hierarchy. Though the idea in \cite{Xiao2005} is
different from that in \cite{Bogdanov2003}, the well-known
Zakharov reduction of the Sato function for the dispersionless KP
(dKP) hierarchy can be also obtained as a result. The critical
step in the procedure of this quasiclassical limit is to find the
asymptotic forms of the potentials $q_i(\frac{T}{\epsilon})$ and
$r_i(\frac{T}{\epsilon})$ appearing in the constraint of the Sato
operator. Starting from the constrained Sato function for the dKP
hierarchy, the constrained dKP hierarchy (cdKPH) and the dKP
hierarchy with self-consistent sources (dKPHSCS) are therefore
constructed. In contrast to the dKP case, the reduction of the
dispersionless mKP hierarchy has been less studied. In this paper,
we will further develop the idea in \cite{Xiao2005} to investigate
the quasiclassical limit of the squared eigenfunction symmetry
constraint of the Sato operator for the mKP hierarchy
\cite{Oevel1998}. As will be shown below, this quasiclassical
limit will lead to a reduction of the Sato function for the
dispersionless mKP hierarchy. In this case, it is also important
for us to find the asymptotic form of $q_i(\frac{T}{\epsilon})$
and $r_i(\frac{T}{\epsilon})$ appearing in the constraint of the
Sato operator for the mKP hierarchy. The constrained dmKP
hierarchy (cdmKPH) will soon be obtained and its commutability can
be proved. We can see that the dmKP hierarchy of
$(2+1)$-dimensions is decomposed into two commutative
$(1+1)$-dimensional hierarchies of hydrodynamical type. The dmKP
hierarchy with self-consistent sources (dmKPHSCS) and its
associated conservation equations will also be obtained according
to our approach. Utilizing the conservation equations, we can
develop the hodograph reduction method
\cite{Kodama1988,Kodama1989} to solve the dmKPHSCS. In this sense,
the dmKPHSCS may be regarded as an integrable generalization of
the dmKP hierarchy.

The paper will be organized as follows. In section 2, we briefly
review some definitions and results about the mKPHSCS. In section
3, we take the quasiclassical limit to the squared eigenfunction
symmetry constraint of the Sato operator for the mKP hierarchy.
The cdmKPH and the dmKPHSCS together with its associated
conservation equations are also obtained. Section 4 is devoted to
presenting some solutions for the dmKPHSCS obtained by means of
the hodograph reduction methodand.

\section{The squared eigenfunction symmetry constraint of the mKP hierarchy
and the mKP hierarchy with self-consistent sources}
\setcounter{equation}{0} \hskip\parindent We first review some
results about the mKP hierarchy with self-consistent sources in
the framework of Sato theory [25-29]. Given a pseudo-differential
operator (PDO) of the form \cite{Konopelchenko1993}
\begin{equation}
\label{1}
    L=\partial+v_0+v_1\partial^{-1}+v_2\partial^{-2}+...,\ \
\end{equation}
where $v_i=v_i(t)$, the mKP hierarchy is defined as
\begin{equation}
\label{2}
    \partial_{t_k}L=[Q_k,L], \ \ k\in \mathbb{N},
\end{equation}
where $Q_k=(L^k)_{\geq 1}$ means the part of order$\geq1$ of
$L^k$. The Lax euqation (\ref{2}) is equivalent to the
compatibility of the following linear equations
\begin{subequations}
\label{3}
\begin{equation}
\label{3.1}
    L\psi_{mKP} = \lambda\psi_{mKP},
\end{equation}
\begin{equation}
\label{3.2}
    \partial_{t_k}\psi_{mKP} = Q_k\psi_{mKP}.
\end{equation}
\end{subequations}
$\psi_{mKP}$ is often called the eigenfunction of the mKP
hierarchy and also satisfies
\begin{equation}
\label{3.3}
    \partial_\lambda\psi_{mKP}=M_{mKP}\psi_{mKP},
\end{equation}
where $M_{mKP}$ is the Orlov operator of the mKP hierarchy. The
adjoint eigenfunction $\psi_{mKP}^*$ satisfies
\begin{subequations}
\label{4}
\begin{equation}
\label{4.1}
    L^*\psi_{mKP}^* = \lambda \psi_{mKP}^*,
\end{equation}
\begin{equation}
\label{4.2}
    \partial_k\psi_{mKP}^* = -Q_k^* \psi_{mKP}^*,
\end{equation}
\begin{equation}
\label{4.3}
    \partial_\lambda\psi_{mKP}^* = -M_{mKP}^* \psi_{mKP}^*.
\end{equation}
\end{subequations}

Making a constraint of the PDO $L$ (\ref{1}) as
\cite{Oevel1998,Xiao20042}
\begin{equation}
\label{5}
    L^n=(L^n)_{\geq 1}+\s q_i(t)\partial^{-1}r_i(t)\p ,\ \ \
    n\in \mathbb{N},
\end{equation}
where $q_i(t)$ and $r_i(t)$ satisfying
\begin{equation}
\label{6}
    q_{i,t_k} = Q_k(q_i),\ \ r_{i,t_k} = -(\p Q_k\p^{-1})^*(r_i)=-(\p^{-1}Q_k^*\p)(r_i),\ \ i=1,...,N,
    k\in\mathbb{N},
\end{equation}
and $Q_k=[(L^n)^{\frac{k}{n}}]_{\geq 1}=[((L^n)_{\geq 1}+\s
q_i(t)\partial^{-1}r_i(t)\p)^{\frac{k}{n}}]_{\geq 1}$, we will get
the ($n$-)constrained mKP hierarchy (cmKPH) as
\begin{subequations}
\label{7}
\begin{equation}
\label{71}
     (L^n)_{t_k}=[Q_k,L^n],
\end{equation}
\begin{equation}
\label{b012}
    q_{i,t_k}=Q_k(q_i),
\end{equation}
\begin{equation}
\label{b013}
    r_{i,t_k}=-(\p^{-1}Q_k^*\p)(r_i), \ \ i=1,...,N.
\end{equation}
\end{subequations}
(\ref{5}) with (\ref{6}) is called the squared eigenfunction
symmetry constraint for the mKP hierarchy \cite{Oevel1998} and
$r_i$ is called the integrated adjoint eigenfuntion, i.e.,
$r_{i,x}$ satisfies
(\ref{4.2}).\\
If we add the term $(Q_k)_{t_n}$ to the right side of equation
(\ref{71}) and requiring $k<n$, the mKP hierarchy with
self-consistent sources (mKPHSCS) will be obtained as
\cite{Xiao20042}
\begin{subequations}
\label{8}
\begin{equation}
\label{81}
     (Q_k)_{t_n}-(L^n)_{t_k}+[Q_k,L^n]=0,
\end{equation}
\begin{equation}
\label{82}
    q_{i,t_k}=Q_k(q_i),
\end{equation}
\begin{equation}
\label{83}
    r_{i,t_k}=-(\p^{-1}Q_k^*\p)(r_i),\ \ \ i=1,...,N.
\end{equation}
\end{subequations}
As the case of KPHSCS \cite{Xiao20041} and many other cases in
(1+1)-dimensions (see \cite{Zeng2000,Zeng2002} and the references
therein), the cmKPH (\ref{7}) may be considered as the stationary
one of the mKPHSCS (\ref{8}) if '$t_n$' is viewed as the evolution
variable. \\
When $k=2$, $n=3$ in (\ref{8}), we will get the mKP
equation with self-consistent sources (mKPESCS) (with $y=t_2$,
$t=t_3$,$v=v_0$)\cite{Xiao20042}
\begin{subequations}
\label{9}
\begin{equation}
\label{91}
     v_t-\frac{1}{4}v_{xxx}-\frac{3}{4}D_x^{-1}(v_{yy})-\frac{3}{2}D_x^{-1}(v_y)v_x+\frac{3}{2}v^2v_x+\sum_{i=1}^N(q_ir_i)_x= 0,
\end{equation}
\begin{equation}
\label{92}
    q_{i,y} = q_{i,xx}+2vq_{i,x},
\end{equation}
\begin{equation}
\label{93}
    r_{i,y} = -r_{i,xx}+2vr_{i,x}, \ \ \ i=1,\cdots,N,
\end{equation}
\end{subequations}
where $D_x^{-1}=\int^x$ is the integral operator. Under (\ref{92})
and (\ref{93}), (\ref{91}) will be obtained by the compatibility
of the following auxiliary linear equations (Lax pair)
\begin{subequations}
\begin{equation}
\psi_y = \psi_{xx}+2v\psi_x,
\end{equation}
\begin{equation}
\psi_t =
\psi_{xxx}+3v\psi_{xx}+\frac{3}{2}[v_x+v^2+D_x^{-1}(v_y)]\psi_x+\sum_{i=1}^Nq_iD_x^{-1}(r_i\psi_x).
\end{equation}
\end{subequations}

\section{The quasiclassical limit}
\setcounter{equation}{0} \hskip\parindent In this section, we will
consider the quasiclassical limit of (\ref{5}) with (\ref{6})
which will give rise to the dispersionless counterpart of
(\ref{7}) and (\ref{8}), i.e. the constrained dispersionless mKP
hierarchy (cdmKPH) and dispersionless mKP hierarchy with
self-consistent sources
(dmKPHSCS).\\
Taking $T_n=\epsilon t_n$ and thinking of
$v_i(\frac{T}{\epsilon})=V_i(T)+O(\epsilon)$ as
$\epsilon\rightarrow 0$, $L$ in (\ref{1}) changes into
\begin{equation}
\label{10}
    L_\epsilon=\epsilon\partial+\sum_{i=0}^\infty
    v_i(\frac{T}{\epsilon})(\epsilon\partial)^{-i}=\epsilon\partial+\sum_{i=0}^\infty
    (V_i(T)+O(\epsilon))(\epsilon\partial)^{-i},\ \ \ \
    \partial=\partial_X,\ \ \ X=\epsilon x.
\end{equation}
The constraint (\ref{5}) now changes into
\begin{equation}
\label{11}
    L^n_\epsilon=Q_{\epsilon n}+\sum_{i=1}^Nq_i(\frac{T}{\epsilon})(\epsilon\partial)^{-1}r_i(\frac{T}{\epsilon})\epsilon\partial,\ \ Q_{\epsilon n}=(L^n_\epsilon)_{\geq
    1},
\end{equation}
where $q_i(\frac{T}{\epsilon})$ and $r_i(\frac{T}{\epsilon})$
satisfy
\begin{equation}
\label{12}
    \epsilon [q_i(\frac{T}{\epsilon})]_{T_k} = Q_{\epsilon
    k}(q_i(\frac{T}{\epsilon})),\ \ \epsilon [r_i(\frac{T}{\epsilon})]_{T_k} = -[(\epsilon \p)^{-1}Q^*_{\epsilon
    k}(\epsilon \p)](r_i(\frac{T}{\epsilon})),\ \ Q_{\epsilon
    k}=[(L^n_\epsilon)^{\frac{k}{n}}]_{\geq 1},\ \ i=1,...,N.
\end{equation}
It is easy to prove that
\begin{equation}
\label{14}
    {\mathcal{L}} = \sigma^{\epsilon}(L_\epsilon) = p+\sum_{i=0}^\infty
    V_i(T)p^{-i}.
\end{equation}
is a solution of the dmKP hierarchy, i.e., satisfies
\begin{equation}
\label{15}
    \partial_{T_n}{\mathcal{L}} = \{{\mathcal{Q}}_n,
    {\mathcal{L}}\},
\end{equation}
where $\sigma^\epsilon$ denotes the principal symbol
\cite{Takasaki1995}, the Poisson bracket is defined as
\begin{equation}
\label{151} \{A(p,x), B(p,x)\} = \frac{\partial A}{\partial p}
\frac{\partial B}{\partial x}-\frac{\partial A}{\partial
x}\frac{\partial B}{\partial p},
\end{equation}
and ${\mathcal{Q}}_n=({\mathcal{L}}^n)_{\geq 1}$ now refers to
powers of $p$ with order $\geq 1$.\\ The dmKP hierarchy can be
also written in the zero-curvature form
\begin{equation}
\label{16}
    \frac{\partial {\mathcal{Q}}_l}{\partial T_s}-\frac{\partial {\mathcal{Q}}_s}{\partial
    T_l}+\{{\mathcal{Q}}_l,{\mathcal{Q}}_s\}=0,\ \ l,s \in \mathbb{N}.
\end{equation}
When $l=2$, $s=3$, we will get the dmKP equation ($Y=T_2$, $T=T_3$
and $V=V_0$ )
\begin{equation}
\label{161}
    2V_T-\frac{3}{2}D_X^{-1}(V_{YY})-3V_XD_X^{-1}(V_Y)+3V^2V_X=0.
\end{equation}
Similarly like the dKP case \cite{Takasaki1995}, from (\ref{3}),
(\ref{3.3}) and (\ref{4}), it can be proved that
$\psi_{mKP}(\frac{T}{\epsilon})$ \cite{Chang2000} and
$\psi_{mKP}^*(\frac{T}{\epsilon})$ has the following WKB
asymptotic expansion as $\epsilon\rightarrow 0$,
\begin{equation}
\label{17}
    \psi_{mKP}(\frac{T}{\epsilon},\lambda)=exp[\frac{1}{\epsilon}S_{mKP}(T,\lambda)+O(1)],\
    \ \psi^*_{mKP}(\frac{T}{\epsilon},\lambda)=exp[-\frac{1}{\epsilon}S_{mKP}(T,\lambda)+O(1)],\ \epsilon\rightarrow 0.
\end{equation}
From (\ref{3.2}) and (\ref{17}), we can obtain a hierarchy of
conservation equations for the momentum function $p=\frac{\partial
S_{mKP}}{\partial X}$,
\begin{equation}
\label{19}
    \frac{\partial p}{\partial T_l}=\frac{\partial {\mathcal{Q}}_l(p)}{\partial
    X},\ \ l\in\mathbb{N},
\end{equation}
the compatibility of which, i.e., $\frac{\partial^2 p}{\partial
T_l \partial T_s}=\frac{\partial^2 p}{\partial T_s \partial T_l}$
implies the dmKP hierarchy (\ref{16}). \\
Since $q_i(t)$ and $r_i(t)_x$ evolve in the same rule as
$\psi_{mKP}$ and $\psi^*_{mKP}$ respectively, we regard
\begin{subequations}
\label{20}
\begin{equation}
\label{201}
     q_i(\frac{T}{\epsilon})=\psi_{mKP}(\frac{T}{\epsilon},\lambda=\lambda_i) \sim
     exp[\frac{S_{mKP}(T,\lambda_i)}{\epsilon}+a_{i1}+O(\epsilon)],
\end{equation}
\begin{equation}
\label{202}
    \epsilon[r_i(\frac{T}{\epsilon})]_X=\psi_{mKP}^*(\frac{T}{\epsilon},\lambda=\lambda_i) \sim
     exp[-\frac{S_{mKP}(T,\lambda_i)}{\epsilon}+a_{i2}+O(\epsilon)],\
     \ \epsilon\rightarrow 0,\ \ i=1,...,N.
\end{equation}
\end{subequations}
By a tedious computation, we will find that when
$\epsilon\rightarrow 0$,
\begin{equation}
\nonumber
\begin{array}{ll}
    q_i(\frac{T}{\epsilon})(\epsilon
    \partial)^{-1}r_i(\frac{T}{\epsilon})(\epsilon
    \p)=&-\frac{e^{a_{i1}+a_{i2}}}{S_{mKP}(T,\la_i)_X}[1+(S_{mKP}(T,\la_i)_X+O(\epsilon))(\epsilon\partial)^{-1}+\cdots\\
    &+(( S_{mKP}(T,\lambda_i)_X)^n+O(\epsilon))(\epsilon\partial)^{-n}+\cdots].
\end{array}
\end{equation}
Taking the principal symbol of both sides of (\ref{11}), we obtain
the constraint for ${\mathcal{L}}$ as
\begin{eqnarray}
\label{21}
{\mathcal{L}}^n&=&{\mathcal{Q}}_n-\sum_{i=1}^N\frac{e^{a_{i1}+a_{i2}}}{S_{mKP}(T,\la_i)_X}[1+S_{mKP}(T,\la_i)_Xp^{-1}+\cdots+(S_{mKP}(T,\la_i)_X)^n p^{-n}+\cdots]\nonumber\\
&=&{\mathcal{Q}}_n-\sum_{i=1}^N(\frac{a_i}{p_i}+\frac{a_i}{p-p_i}),
\end{eqnarray}
where ${\mathcal{Q}}_n=({\mathcal{L}}^n)_{\geq 1}$ and
\begin{equation}
\label{22} a_i=e^{a_{i1}+a_{i2}},\ \ p_i=S_{mKP}(T,\lambda_i)_X.
\end{equation}
From (\ref{12}), (\ref{20}) and (\ref{22}), a direct tedious
computation shows the evolution for $a_i$ and $p_i$, i.e.,
\begin{subequations}
\label{23}
\begin{equation}
\label{231}
 a_{i,T_k}=[a_i(\frac{\partial}{\partial p}{\mathcal{Q}}_k(p))|_{p=p_i}]_X,
\end{equation}
\begin{equation}
\label{232}
   p_{i,T_k}=[{\mathcal{Q}}_k(p)|_{p=p_i}]_X,\ \ k\in{\mathbb{N}}
    i=1,\cdots,N,
\end{equation}
\end{subequations}
where
\begin{equation}
\label{233}
{\mathcal{Q}}_k=[({\mathcal{L}}^n)^{\frac{k}{n}}]_{\geq
1}=\{[{\mathcal{Q}}^n-\s(\frac{a_i}{p_i}+\frac{a_i}{p-p_i})]^{\frac{k}{n}}\}_{\geq
1}.
\end{equation}
It is easy to prove that with (\ref{23}), the constraint
(\ref{21}) is compatible with the dmKP hierarchy (\ref{15}). So
the quasiclassical limit of the constrained mKP hierarchy
(\ref{7}) gives rise to the following constrained dispersionless
mKP hierarchy (cdmKPH)
\begin{subequations}
\label{24}
\begin{equation}
\label{241}
     ({\mathcal{L}}^n)_{T_k} = \{{\mathcal{Q}}_k,{\mathcal{L}}^n\},
\end{equation}
\begin{equation}
\label{242}
     a_{i,T_k}=[a_i(\frac{\partial}{\partial p}{\mathcal{Q}}_k(p))|_{p=p_i}]_X,
\end{equation}
\begin{equation}
\label{243}
    p_{i,T_k}=[{\mathcal{Q}}_k(p)|_{p=p_i}]_X,\ \
    i=1,...,N,
\end{equation}
\end{subequations}
where ${\mathcal{L}}^n$ and ${\mathcal{Q}}_k$ are given by
(\ref{21}) and (\ref{233}) respectively.
\begin{theorem}
The equations with different $k$ in (\ref{24}) are commutative.
\end{theorem}
{\bf{Proof}}.From (\ref{241}), we can prove that for $\forall$
$k_1$,$k_2$$\in {\mathbb{N}}$, the following identity holds,
\begin{equation}
\label{244}
(Q_{k_1})_{T_{k_2}}-(Q_{k_2})_{T_{k_1}}+\{Q_{k_1},Q_{k_2}\}=0,
\end{equation}
where $Q_{k_i}=[(L^n)^{\frac{k_i}{n}}]_{\geq 1}$, $i=1,2$. A
direct computation shows that
$$(p_i)_{T_{k_1}T_{k_2}}=\{[Q_{k_1}(p)|_{p=p_i}]_X\}_{T_{k_2}}=[(Q_{k_1})_{T_{k_2}}|_{p=p_i}+(Q_{k_1})_p|_{p=p_i}((Q_{k_2})_X|_{p=p_i}+(Q_{k_2})_p|_{p=p_i}p_{i,X}))]_X.$$
So, $(p_i)_{T_{k_1}T_{k_2}}=(p_i)_{T_{k_2}T_{k_1}}$ is equivalent
to
$$(Q_{k_1})_{T_{k_2}}|_{p=p_i}+(Q_{k_1})_p(Q_{k_2})_X|_{p=p_i}=(Q_{k_2})_{T_{k_1}}|_{p=p_i}+(Q_{k_2})_p(Q_{k_1})_X|_{p=p_i},$$
which holds from (\ref{244}).\\
Analogously, we can prove
$(a_i)_{T_{k_1}T_{k_2}}=(a_i)_{T_{k_2}T_{k_1}}$. The proof for
$({\mathcal{L}}^n)_{T_{k_1}T_{k_2}}=({\mathcal{L}}^n)_{T_{k_2}T_{k_1}}$
is similar to that for the mKP hierarchy.\\This completes the proof.\\
In fact, the commutativity for the cmKPH (\ref{24}) is a result of
that for the mKP hierarchy since the constraint (\ref{21}) and
(\ref{23}) is compatible with the mKP hierarchy.\\
As will be shown below, the $(2+1)$-dimensional dispersionless mKP
hierarchy (\ref{16}) is decomposed into two commutative
$(1+1)$-dimensional hierarchies of hydrodynamic type of (\ref{24})
with $k=s$ and $k=l$ respectively. For example, when $n=1$,
($\ref{21}$) becomes
\begin{equation}
\label{24.1}
{\mathcal{L}}=p-\sum_{i=1}^N(\frac{a_i}{p_i}+\frac{a_i}{p-p_i}),
\end{equation}
and we immediately get the following two systems of hydrodynamic
type when $k=2$ and $k=3$ ($Y=T_2$, $T=T_3$),
\begin{subequations}
\label{24.2}
\begin{equation}
\label{24.21} a_{i,Y}=2[a_i(p_i-S_1)]_X,
\end{equation}
\begin{equation}
\label{24.22} p_{i,Y}=[p_i^2-2S_1p_i]_X,
\end{equation}
\end{subequations}
and
\begin{subequations}
\label{24.3}
\begin{equation}
\label{24.31} a_{i,T}=[a_i(3p_i^2-6S_1p_i+3S_1^2-3S_2)]_X,
\end{equation}
\begin{equation}
\label{24.32} p_{i,T}=[p_i^3-3S_1p_i^2+(3S_1^2-3S_2)p_i]_X,
\end{equation}
\end{subequations}
where $S_1=\s \frac{a_i}{p_i}$ and $S_2=\s a_i$. It is easy to
verify that common solution of the systems (\ref{24.2}) and
(\ref{24.3}) generates a solution $V=-\s \frac{a_i}{p_i}$ of the
dmKP equation (\ref{161}). Generally, if
${\mathcal{L}}^n={\mathcal{Q}}_n-\sum_{i=1}^N(\frac{a_i}{p_i}+\frac{a_i}{p-p_i})$
satisfies (\ref{24}) with $k=s$ and $k=l$ simultaneously, then
${\mathcal{Q}}_i=[({\mathcal{L}}^n)^{\frac{i}{n}}]_{}\geq 1$ with
$i=l,s$ will satisfy (\ref{16}). \\
When adding the term $({\mathcal{Q}}_k)_{T_n}$ to the right hand
side of (\ref{241}) and requiring $k<n$, or taking the
quasiclassical limit of (\ref{8}) directly, we will obtain the
dmKP hierarchy with self-consistent sources (dmKPHSCS) as
\begin{subequations}
\label{25}
\begin{equation}
\label{251}
     ({\mathcal{Q}}_k)_{T_n}-({\mathcal{L}}^n)_{T_k}+ \{{\mathcal{Q}}_k,{\mathcal{L}}^n\}=0,
\end{equation}
\begin{equation}
\label{252}
     a_{i,T_k}=[a_i(\frac{\partial}{\partial p}{\mathcal{Q}}_k(p))|_{p=p_i}]_X,
\end{equation}
\begin{equation}
\label{253}
    p_{i,T_k}=[{\mathcal{Q}}_k(p)|_{p=p_i}]_X,\ \
    i=1,...,N,
\end{equation}
\end{subequations}
where ${\mathcal{L}}^n$ and ${\mathcal{Q}}_k$ are given by
(\ref{21}) and (\ref{233}) respectively. If "$T_n$" is viewed as
the evolution variable, the cdmKPH (\ref{24}) may be regarded as
the stationary system of the dmKPHSCS (\ref{25}) like the
dispersionfull case. It is not difficult to prove that under
(\ref{252}) and (\ref{253}), the equation (\ref{251}) will be
obtained by the compatibility of the following conservation
equations
\begin{subequations}
\label{26}
\begin{equation}
\label{261}
     p_{T_k}=[{\mathcal{Q}}_k(p)]_X,
\end{equation}
\begin{equation}
\label{262}
     p_{T_n}=[{\mathcal{L}}^n(p)]_X=[{\mathcal{Q}}_n(p)-\sum_{i=1}^N(\frac{a_i}{p_i}+\frac{a_i}{p-p_i})]_X.
\end{equation}
\end{subequations}
For example, when $k=2$, $n=3$,
\begin{equation}
\nonumber
     {\mathcal{L}}^3=p^3+3V_0p^2+(3V_0^2+3V_1)p-\sum_{i=1}^N(\frac{a_i}{p_i}+\frac{a_i}{p-p_i}),
\end{equation}
(\ref{25}) becomes the dmKP equation with self-consistent sources
(dmKPESCS) ($Y=T_2$, $T=T_3$, $V=V_0$ and $V_1$ is eliminated by
$V_{1,X}=\frac{1}{2}V_{0,Y}-\frac{1}{2}(V_0^2)_X$)
\begin{subequations}
\label{27}
\begin{equation}
\label{271}
     2V_T-\frac{3}{2}D_X^{-1}(V_{YY})-3V_XD_X^{-1}(V_Y)+3V^2V_X-2\s(\frac{a_i}{p_i})_X=0,
\end{equation}
\begin{equation}
\label{272}
     a_{i,Y}=2[a_i(p_i+V)]_X,
\end{equation}
\begin{equation}
\label{273}
    p_{i,Y}=(p_i^2+2Vp_i)_X,\ \ i=1,...,N.
\end{equation}
\end{subequations}
Under (\ref{272}) and (\ref{273}), (\ref{271}) will be obtained by
the compatibility of the following conservation equations
\begin{subequations}
\label{28}
\begin{equation}
\label{281}
     p_Y=2pp_X+2V_Xp+2Vp_X,
\end{equation}
\begin{equation}
\label{282}
     p_T=3V_Xp^2+(3V^2+V_1)_Xp+(3p^2+6Vp+3V^2+V_1)p_X-\sum_{i=1}^N[(\frac{a_i}{p_i})_X+\frac{a_{i,X}}{p-p_i}+\frac{a_i(p_{i,X}-p_X)}{(p-p_i)^2}],
\end{equation}
\end{subequations}
where $V_{1,X}=\frac{3}{2}V_Y-\frac{3}{2}(V^2)_X$.

As will be shown in the next section, utilizing the conservation
equations (\ref{26}), (\ref{25}) will be solved by the hodograph
reduction method. In this sense, the dmKPHSCS (\ref{25}) can be
regarded as an integrable generalization of the dmKPH (\ref{16}).
\section{Hodograph reduction solutions} \setcounter{equation}{0} \hskip\parindent
In this section we will present some solutions of the dmKPHSCS
(\ref{25}) obtained by means of the hodograph reduction method
\cite{Kodama1988,Xiao2005}. Provided the conservation equations
(\ref{26}), one can consider the $M$-reductions of (\ref{26}) so
that the momentum function $p$, the auxiliary potentials $V_i$,
$i\geq 1$ and $a_i$, $p_i$,$i=1,...,N$ depend only on a set of
functions $W=$($W_1$,...,$W_M$) with $W_1=V_0=V$ and
($W_1$,...,$W_M$) satisfy commuting flows
\begin{equation}
\label{51} \frac{\partial W}{\partial T_n} = A_n(W)\frac{\partial
W}{\partial X},\ \ n\geq 2
\end{equation}
where the $M\times M$ matrices $A_n$ are functions of
$(W_1,...,W_M)$ only. In the following, for example, we list some
results for the dmKPESCS (\ref{27}) in the cases of $M=1$ and
$M=2$.
1. $M=1$\\
In this case, we will get the following hodograph equation
\begin{equation}
\label{58}
      X+(\frac{4c-4}{c-2}V)Y+[\frac{3}{4}(\frac{4c-4}{c-2})^2V^2+\frac{3c}{2(c-2)}V^2+b+(\sum_{i=1}^Nd_i)V^{\frac{4-c}{c-2}}]T=F(V),
\end{equation}
where $c$,$d_i$ are constants and $F(V)$ is an arbitrary function
of
$V$.\\
(1)$c=3$, $\frac{4-c}{c-2}=1$, $F(V)=0$. Then
\begin{equation}
\label{59} V=\frac{-[(\s d_i)T+8Y]\pm \sqrt{[(\s
d_i)T+8Y]^2-210(bT+X)}}{105T},
\end{equation}
and $V$, $p_i=2V$, $a_i=d_iV^3$, $i=1,\cdots,N$ is an explicit
solution for the dmKPESCS (\ref{27}). \\
(2)$c=\frac{8}{3}$, $\frac{4-c}{c-2}=2$, Choosing $F(V)=0$. Then
\begin{equation}
\label{510} V=\frac{-5Y\pm \sqrt{25Y^2-(\s d_i+81)(bT+X)T}}{(\s
d_i+81)T},
\end{equation}
and $V$, $p_i=3V$, $a_i=d_iV^4$, $i=1,\cdots,N$ is another
solution for
the dmKPESCS (\ref{27}).\\
{\bf{Remark}}:When $\s d_i=0$, (\ref{59})) and (\ref{510})) will
degenerate to the solution for the dmKP equation (\ref{161}).\\
2. $M=2$\\
In this case we can get the following solution of (\ref{27}) as
\begin{subequations}
\label{522}
\begin{equation}
\label{5221} V=\frac{-Y\pm\sqrt{-Y^2+2(3T+2)[(\s
c_i)T+X]}}{2(3T+2)},
\end{equation}
\begin{equation}
\label{5222} a_i=-c_iV(V+\frac{Y}{3T+2}),
\end{equation}
\begin{equation}
\label{5223} p_i=-\frac{Y}{3T+2}-V,\ \ i=1,...,N.
\end{equation}
\end{subequations}
with $c_i$ are constants.\\
{\bf{Remark}}: (\ref{5221})
degenerates to the solution of the dmKP equation (\ref{161}) when
$\sum_{i=1}^Nc_i=0$.

\section*{Acknowledgment}\hskip\parindent
This work was supported by the Chinese Basic Research Project
"Nonlinear Science".

\hskip\parindent


\begin{thebibliography}{s99}
\bibitem{Kuperschmidt1977}
B.A.Kuperschmidt and Yu.I.Manin, Func.Anal.Appl.I 11(3) (1977)
31-42; II 17(1) (1978) 25-37.

\bibitem{Lebedev1979}
D.Lebedev and Yu.I.Manin, Phys.Lett. 74A(1979) 154-156.

\bibitem{Zakharov1980}
V.E.Zakharov, Func.Anal.Priloz. 14 (1980) 89-98; Physica 3D (1981)
193-202.

\bibitem{Lax1983}
P.D.Lax and C.D. Levermore, Commun.Pure Appl.Math. 36 (1983)
253-290, 571-593, 809-830.

\bibitem{Krichever1988}
I.M.Krichever, Func.Anal.Priloz. 22 (1988) 37-52.

\bibitem{Kodama1988}
Y.Kodama, Phys.Lett. 129A (1988) 223-226; 147A (1990) 477-482.

\bibitem{Kodama1989}
Y.Kodama and J.Gibbons, Phys.Lett.A 135 II (1989) 167-170.

\bibitem{Takasaki1992}
K.Takasaki and T.Takebe, Int.J.Mod.Phys.A Suppl.1B (1992) 889-922.

\bibitem{Takasaki1995}
K.Takasaki and T.Takebe, Rev.Math.Phys.7 (1995) 743-808.

\bibitem{Zakharov1994}
V.E.Zakharov, in: Singular limit of dispersive waves, ed.
N.M.Erconali et al (Plenum, New York, 1994) 165-174.

\bibitem{Krichever1992}
I.M.Krichever, Commun.Math.Phys. 143 (1992) 415-429.

\bibitem{Aoyama1996}
S.Aoyama and Y.Kodama, Commun.Math.Phys. 182 (1996) 185-219.

\bibitem{Konopelchenko2001}
B.Konopelchenko, L.Martinez Alonso and O. Ragnisco, J.Phys.A 34
(2001) 10209-10217.

\bibitem{Konopelchenko2002}
B.Konopelchenko, L.Martinez Alonso and E.Medina,
arXiv:nlin.SI/0202013.

\bibitem{Bogdanov2001}
L.V.Bogdanov, B.G.Konopelchenko and L.Martinez Alonso,
arXiv:nlin.SI/0111062.

\bibitem{Bogdanov2003}
L.V.Bogdanov and B.G.Konopelchenko, arXiv:nlin.SI/0312013.

\bibitem{Martinez2003}
L.Martinez Alonso and M.Man\~{a}s, arXiv:nlin.SI/0302009.

\bibitem{Manas2002}
M.Man\~{a}s, L.Martinez Alonso and E.Medina, J.Phys.A 35 (2002)
401-417.

\bibitem{Mel'nikov1987}
V.K.Mel'nikov, Commun.Math.Phys. 112 (1987) 639-652.

\bibitem{Mel'nikov1989}
V.K.Mel'nikov, Commun.Math.Phys. 126 (1989) 201-215.

\bibitem{Leon1990}
J.Leon and A.Latifi, J.Phys.A 23 (1990) 1385-1403.

\bibitem{Zeng2000}
Yunbo Zeng, Wenxiu Ma and Runliang Lin, J.Math.Phys. 41 (2000) 8.

\bibitem{Zeng2002}
Yunbo Zeng, Wenxiu Ma and Yijun Shao, J.Math.Phys. 42(5) (2001)
2113-2128.

\bibitem{Zeng2003}
Yunbo Zeng, Yijun Shao and Weimin Xue, J.Phys.A 36 (2003)
5035-5043.

\bibitem{Xiao2005}
Ting Xiao and Yunbo Zeng, submitted to Letters in Mathematical
Physics.

\bibitem{Xiao20041}
Ting Xiao and Yunbo Zeng, J.Phys.A 37 (2004) 7143-7162.

\bibitem{Xiao20042}
Ting Xiao and Yunbo Zeng, Physica A 353 (2005) 38-60.

\bibitem{Sato1980}
M.Sato, RIMS Kokyuroku (Kyoto Univ.) 439 (1980) 30.

\bibitem{Jimbo1983}
L.A.Dickey, Soliton equation and Hamiltonian systems (Singapore,
World Scientific, 1991).

\bibitem{Konopelchenko1993}
B.Konopelchenko and W.Oevel, Publ.RIMS, Kyoto Univ. 29 (1993) 581.

\bibitem{Oevel1998}
W.Oevel and S. Carillo, J. Math. Anal. Appl. 217 (1998) 161-178.

\bibitem{Chang2000}
Jen-Hsu Chang and Ming-Hsien Tu, J. Math. Phys. 41 (2000)
5391-5406.

\end {thebibliography}
\end{document}